\documentstyle[multicol,aps,epsf]{revtex}
\def\d{{\rm d}}
\begin{document}
\title{The Mechanical Response of Vacuum}
\author{Ramin Golestanian}
\address{Institute for Advanced Studies in Basic Sciences,
Zanjan   45195-159, Iran}
\author{Mehran Kardar}
\address{Department of Physics, Massachusetts Institute of  
Technology, Cambridge, MA 02139}
\date{\today}
\maketitle
\begin{abstract}
A path integral formulation is developed for the {\it dynamic}  
Casimir effect. It allows us to study arbitrary deformations in
{\it space and time} of the perfectly reflecting (conducting) 
boundaries of a cavity. The mechanical response of the 
intervening vacuum is calculated to linear order in the 
frequency--wavevector plane. For a single corrugated 
plate we find a correction to mass at low frequencies, 
and an effective shear viscosity at high frequencies; 
both anisotropic. For two plates there is  resonant 
dissipation for {\it all frequencies} greater than the lowest 
optical mode of the cavity.
\end{abstract}
\pacs{03.70.+k, 42.50.Ct, 11.10.-z, 78.60.Mq 
}
\begin{multicols}{2}
The standard Casimir effect\cite{Casimir} is a macroscopic  
manifestation of quantum fluctuations of vacuum. 
The modified boundary conditions of the electromagnetic field
in the space between two parallel conducting plates change zero  
point vacuum fluctuations, resulting in an attractive force between 
the plates, which is within the sensitivity of  current experimental 
force apparatus\cite{CasX}. Thus, by measuring the mechanical 
force between macroscopic bodies, it is in principle possible to
gain information about the behavior of the quantum vacuum.
Although less well known than its static counterpart, the
dynamical Casimir effect, describing the force and radiation
from moving mirrors has also garnered much attention
\cite{Moore,Fulling,Jaekel,Neto,Cavity,Meplan,Lambrecht}.
This is partly due to connections to Hawking and Unruh effects
(radiation from black holes and accelerating bodies, respectively),
suggesting a deeper link between quantum mechanics,
relativity, and cosmology\cite{Davis}.

The creation of photons by moving mirrors was first obtained 
by Moore\cite{Moore} for a 1 dimensional cavity.  
Fulling and Davis\cite{Fulling} demonstrated that there is a
corresponding  force, even for a single mirror,
which depends on the third time derivative of its displacement.
These computations take advantage of conformal symmetries of the
1+1 dimensional space time, and can not be easily generalized to 
higher dimension. Furthermore, the calculated force has causality 
problems reminiscent of the radiation reaction forces in classical 
electron theory \cite{Jaekel}. It has been shown that this problem 
is an artifact of the unphysical assumption of perfect reflectivity
of the mirror, and can be resolved by considering realistic 
frequency dependent reflection and transmission from the 
mirror \cite{Jaekel}. 

Another approach to the problem starts with the fluctuations in the
force on a single plate. The fluctuation--dissipation theorem is then
used to obtain the mechanical response function\cite{Neto},
whose imaginary part is related to the dissipation. This method
does not have any causality problems, and can also be extended
to higher dimensions. (The force in 1+3 dimensional space-time
depends on the fifth power of the motional frequency.)
The emission of photons by a perfect cavity, and the observability
of this energy, has been studied by different approaches
\cite{Cavity,Meplan,Lambrecht}. The most promising candidate is
the  resonant production of photons when the mirrors
vibrate at the optical resonance frequency of the cavity\cite{Davis}. 
A review, and more extensive references are found in Ref.\cite{Barton}.
More recently, the radiation due to vacuum fluctuations of a 
collapsing bubble has been proposed as a possible explanation 
for the intriguing phenomenon of sonoluminescense
\cite{Eberlein,Knight}. 

In this letter we present a path integral formulation, applicable to all 
dimensions,  for the  problem of perfectly reflecting 
mirrors that undergo arbitrary dynamic deformations\cite{Ford}. 
We calculate the frequency--wavevector dependent mechanical 
response function, defined as the ratio between the induced force 
and the deformation field, in the linear regime. 
From the response function we extract a plethora of interesting
results, some of which we list here for the specific example
of lateral vibrations of uniaxially corrugated plates:
{\bf (1)} A single plate with corrugations of wavenumber ${\bf k}$,
vibrating at frequencies $\omega\ll ck$, obtains  {\it anisotropic}
corrections to its mass.
{\bf (2)} For $\omega\gg ck$, there is dissipation
due to a frequency dependent anisotropic shear viscosity. 
{\bf (3)} A second plate at a separation $H$ modifies
the mass renormalization by a function of $kH$, but does not change
the dissipation for frequencies $\omega^2<(ck)^2+(\pi c/H)^2$.
{\bf (4)} For all frequencies higher than this first optical normal mode
of the cavity, the mechanical response is infinite, implying 
that such modes can not be excited by any finite external force.
This is intimately connected to the resonant particle creation 
reported in the  
literature\cite{Cavity,Meplan,Lambrecht,Davis}. 
{\bf (5)} A phase angle $\theta$ between two similarly 
corrugated parallel plates results in Josephson--like effects: 
There is a static force proportional to $\sin (\theta)$, while a
uniform relative velocity results in an oscillating force.
{\bf (6)} We calculate a (minute) correction to the velocity
of capillary waves on the surface of mercury due to a small
change in its surface tension.

Our approach is a natural generalization of the path integral methods  
developed by Li and Kardar\cite{LiK} to study fluctuation-induced 
interactions between deformed manifolds embedded in a correlated 
fluid. Such interactions result from {\it thermal fluctuations} of the fluid.
These methods are readily generalized to zero point quantum  
fluctuations of a field, taking advantage of the path integral 
quantization formalism. Since in the Euclidian path integral 
formulation the space and time coordinates are equivalent, 
deformations of the boundaries in space and time appear on the
same footing. As is usual, we simplify the problem by considering
a scalar field $\phi$ (in place of the electromagnetic vector
potential\cite{EM}) with the action
\begin{equation}
S=\frac{1}{2} \int \d^d X \;\partial_{\mu}\phi(X)
			\partial_{\mu}\phi(X),\label{action}
\end{equation}
where summation over $\mu=1,\cdots,d$ is implicit. 
Following a Wick rotation, imaginary time appears as another
coordinate $X_d=ict$ in the $d$--dimensional space-time. 
We would like to quantize the field subject to the constraints 
of its vanishing on a set of $n$ manifolds (objects) defined
by $X=X_{\alpha}(y_{\alpha})$, where $y_{\alpha}$ parametrize 
the $\alpha$th manifold. Following Ref.\cite{LiK}, we implement 
the constraints using delta functions, and write the partition 
function as
\begin{eqnarray}
{\cal Z}&=&\int {\cal D}\phi(X)
		\prod_{\alpha=1}^{n} \prod_{y_{\alpha}}
		\delta\left(\phi\left(X_{\alpha}(y_{\alpha})
		\right)\right)\;\exp\left\{-\frac{1}{\hbar}S[\phi] 
		\right\}.\label{Z1} 
\end{eqnarray}

The delta functions are next represented by integrals over Lagrange  
multiplier fields. Performing the Gaussian integrations over $\phi(X)$ 
then leads to an effective action for the Lagrange multipliers 
which is again Gaussian \cite{LiK}. Evaluating ${\cal Z}$ is thus 
reduced  to calculating the logarithm of the determinant of a kernel. 
Since the Lagrange multipliers are defined on a set of manifolds
with nontrivial geometry, this calculation is generally complicated. 
To be specific, we focus on  two parallel 2d plates 
embedded in 3+1 space-time, and separated by an 
average distance $H$ along the $x_3$-direction. Deformations of the
plates are parametrized by the height functions $h_1({\bf x},t)$ and  
$h_2({\bf x},t)$, where ${\bf x}\equiv(x_1,x_2)$ denotes the two lateral 
space coordinates while $t$ is the time variable. Following 
Ref.\cite{LiK}, $\ln {\cal Z}$ is calculated by a perturbative series in
powers of the height functions. The resulting expression for the 
effective action (in real time), defined by 
$S_{\rm eff}\equiv -i \hbar \ln {\cal Z}$, after eliminating $h$
independent terms, is
\end{multicols}
\begin{eqnarray}
S_{\rm eff}=\frac{\hbar c}{2} \int\frac{\d \omega \d^2 {\bf  
q}}{(2\pi)^3}\;
\left[A_{+}(q,\omega)\left(|h_1({\bf q},\omega)|^2+|h_2({\bf  
q},\omega)|^2\right) 
\hskip 5 cm \right. \label{Seff} \\ 
\left. \hskip 2 cm -A_{-}(q,\omega)\left(h_1({\bf q},\omega)
h_2(-{\bf q},-\omega)+h_1(-{\bf q},-\omega)h_2({\bf q},\omega)\right)  
\right]
+O(h^3).\nonumber
\end{eqnarray}
\begin{multicols}{2}
The kernels $A_{\pm}(q,\omega)$, that are closely related to the  
mechanical response of the system (see below), are functions of 
the separation $H$, but depend on ${\bf q}$ and $\omega$ only 
through the combination $Q^2=q^2-\omega^2/c^2$. The closed 
forms for these kernels involve cumbersome integrals, and are 
not very illuminating.  Instead of exhibiting these formulas, 
we shall describe their behavior in various regions of the 
parameter space. In the limit 
$H \rightarrow \infty$, $A_{-}^{\infty}(q,\omega)=0$, and
\begin{equation}
A_{+}^{\infty}(q,\omega)=\left\{
	\begin{array}{cl}
	- \frac{1}{360 \pi^2 c^5}(c^2 q^2-\omega^2)^{5/2}
	& \mbox{for $\omega < c q$} , \\
	 \\
	i\frac{{\rm sgn}(\omega)}{360 \pi^2 c^5} (\omega^2- c^2  
q^2)^{5/2}
	& \mbox{for $\omega > c q$}, 
	\end{array} \right. \label{A+infty}
\end{equation}
where ${\rm sgn}(\omega)$ is the sign function. 
While the effective action is real for $Q^2>0$, it becomes purely
imaginary for $Q^2<0$.
The latter signifies dissipation of energy \cite{Neto}, presumably  
by generation of photons \cite{Lambrecht}.  It agrees precisely 
with the results obtained previously\cite{Neto} for the special 
case of flat mirrors (${\bf q}=0$).
(Note that dissipation is already present for a single mirror.)

In the presence of a second plate (i.e. for finite $H$), the 
parameter space of the kernels subdivides into three different 
regions as depicted in Fig.~1. In region I ($Q^2>0$ for any $H$), 
the kernels are finite and real, and hence there is no dissipation.
In region IIa where $-\pi^2/H^2 \leq Q^2 <0$, the 
$H$-independent part of $A_{+}$ is imaginary, while the 
$H$-dependent parts of both kernels are real and finite.
(This is also the case at the boundary $Q^2=-\pi^2/H^2$.)
The dissipation in this regime is simply the sum of what would 
have been observed if the individual plates were decoupled, 
and unrelated to the separation $H$. 
By contrast, in region IIb where $Q^2 < -\pi^2/H^2$, both 
kernels diverge with infinite real and imaginary parts\cite{cutoff}.
This $H$-dependent divergence extends all the way to the negative
$Q^2$ axis, where it is switched off by a $1/H^5$ prefactor.

As a concrete example, let us examine the lateral vibration of  
plates with fixed roughness, such as two corrugated mirrors.
The motion of the plates enters through the time dependences 
$h_1({\bf x},t)=h_1({\bf x}-{\bf r}(t))$ and 
$h_2({\bf x},t)=h_2({\bf x})$; i.e. the first plate undergoes lateral  
motion described by ${\bf r}(t)$, while the second plate is stationary.
The lateral force exerted on the first plate is obtained from 
$f_i(t)=\delta S_{\rm eff}/\delta r_i(t)$. Within linear response, 
it is given by 
\begin{equation}
f_i(\omega)=
\chi_{ij}(\omega)\;r_j(\omega)+f_i^{0}(\omega),\label{rough1}
\end{equation}
where the ``mechanical response tensor'' is 
\end{multicols}
\begin{equation}
\chi_{ij}(\omega)=\hbar c \int \frac{\d^2 q}{(2\pi)^2}
\;q_i q_j \left\{\left[A_{+}(q,\omega)-A_{+}(q,0)\right]\;|h_1({\bf  
q})|^2 
+\frac{1}{2} A_{-}(q,0) \left(h_1({\bf q})h_2(-{\bf q})+h_1(-{\bf q})
h_2({\bf q})\right) \right\}, \label{rough}
\end{equation}
and there is a residual force
\begin{equation}
f_i^{0}(\omega)=-\frac{\hbar c}{2}\;2\pi\delta(\omega)\; 
\int \frac{\d^2 q}{(2\pi)^2}
\;i q_i A_{-}(q,0) \left(h_1({\bf q})h_2(-{\bf q})
-h_1(-{\bf q})h_2({\bf q})\right). \label{fi0w}
\end{equation}
\begin{multicols}{2}

For a single corrugated plate with a deformation 
$h({\bf x})=d \cos{{\bf k}\cdot {\bf x}}$, we can easily calculate
the response tensor using the explicit formulas in Eq.(\ref{A+infty}). 
In the limit of $\omega\ll ck$, expanding the result in powers of  
$\omega$ gives $\chi_{ij}=\delta m_{ij}\omega^2+O(\omega^4)$,
where
$\delta m_{ij}=A\hbar d^2 k^3 {\bf k}_i  {\bf k}_j/(288\pi^2 c)$,
can be regarded as corrections to the mass of the plate.
(Cut-off dependent mass corrections also appear, as in Ref.
\cite{Barton}.) 
Note that these mass corrections are {\it anisotropic} with 
$\delta m_{\parallel}=A\hbar  k^5 d^2/(288\pi^2 c)$ and
$\delta m_{\perp}=0$. Parallel and perpendicular components are
defined with respect to ${\bf k}$, and $A$ denotes the area of the
plates. 
The mass correction is inherently very small: For a macroscopic  
sample with $d\approx \lambda=2\pi/k \approx 1$mm, density 
$\approx 15 {\rm gr/cm}^3$, and thickness $t\approx 1$ mm, we find 
$\delta m/m \sim 10^{-34}$. Even for deformations of a microscopic  
sample of atomic dimensions (close to the limits of the applicability 
of our continuum representations of the boundaries),  $\delta m/m$
can only be reduced to around $10^{-10}$. While the actual 
changes in mass are immeasurably small, the hope is that its 
{\it anisotropy} may be more accessible, say by comparing 
oscillation frequencies of a plate in two orthogonal directions.

For $\omega\gg ck$ the response function is imaginary, and we define  
a frequency dependent effective shear viscosity by 
$\chi_{ij}(\omega)=-i\omega\eta_{ij}(\omega)$. This viscosity is also
anisotropic, with $\eta_{\parallel}(\omega)=\hbar A k^2 d^2  
\omega^4/(720\pi^2 c^4)$, and $\eta_{\perp}(\omega)=0$.
Note that the dissipation is proportional to the fifth time derivative 
of displacement, and there is no dissipation for a uniformly 
accelerating plate. However, a freely oscillating plate will undergo 
a damping of its motion. The characteristic decay time for a plate of
mass $M$ is $\tau\approx 2M/\eta$. For the macroscopic plate of 
the previous paragraph, vibrating at a frequency of 
$\omega\approx 2ck$ (in the $10^{12}$Hz range), the decay time is  
enormous, $\tau\sim 10^{18}$s. However, since the decay time 
scales as the fifth power of the dimension, it can be reduced to 
$10^{-12}$s, for plates of order of 10 atoms. However, the required 
frequencies in this case (in the $10^{18}$Hz range) are very large.
Also note that for the linearized forms to remain valid in this high
frequency regime, we must require very small amplitudes, 
so that the typical velocities involved $v\sim r_0\omega$, are 
smaller than the speed of light. These difficulties can be somewhat 
overcome by considering resonant dissipation in the presence of
a second plate.

With two plates at an average distance $H$, the results are  
qualitatively the same for frequencies less than the natural 
resonance of the resulting cavity. There is a renormalization of mass
in region I, and dissipation appears in region IIa, of Fig.~1. 
However, the mass renormalization at low frequencies ($\omega\ll ck$)
is now a function of both $k$ and $H$, with a crossover from the
single plate behavior for $k H \sim 1$. In the limit of $k H \ll 1$,
we obtain $\delta m_{\parallel}=\hbar A B k^2 d^2/48 c H^3$ and 
$\delta m_{\perp}=0$, with $B=-0.453$. Compared to the single plate,
there is an enhancement by a factor of $(kH)^{-3}$ in 
$\delta m_\parallel$. The effective dissipation in region IIa is simply
the sum of those due to individual plates, and contains no $H$
dependence.

There are additional interesting phenomena resulting from resonances.
We find that both real and imaginary parts of $A_{\pm}(q,\omega)$, 
diverge for $\omega^2/c^2 > q^2+\pi^2/H^2$. 
In the example of corrugated plates, we replace $q$ by $k$ to obtain  
a continuous spectrum of frequencies with diverging dissipation. 
Related effects have been reported in the literature for 1+1 
dimensions\cite{Cavity,Meplan,Lambrecht,Davis}, but occuring at 
a {\it discrete} set of frequencies $\omega_n=n \pi c/H$ with 
integer $n\geq2$. These resonances occur when the frequency of the
external perturbation matches the natural normal modes of the cavity,
thus exciting quanta of such modes. In one space dimension,
such modes are characterized by a discrete set of wavevectors
that are integer multiples of $\pi/H$. The restriction to $n\geq2$ is a
consequence of quantum  electrodynamics being a `free' theory 
(quadratic action): only two-photon states can be excited subject to
conservation of energy. Thus the sum of the frequencies of the two 
photons should add up to the external frequency\cite{Lambrecht}. 

In higher dimensions, the appropriate parameter is the combination 
$\omega^2/c^2-q^2$. From the perspective of the excited photons,  
conservation of momentum requires that their two momenta add up 
to $q$, while energy conservation restricts the sum of their frequencies
to $\omega$. The in-plane momentum $q$, introduce a continuous
degree of freedom: the resonance condition can now be satisfied for a
continuous spectrum, in analogy with optical resonators. 
In Ref.\cite{Lambrecht}, the lowest resonance frequency is found to 
be $2\pi c/H$ which seems to contradict our prediction. However, 
the absence of $\omega_1=\pi c/H$ in 1+1 D is due to a vanishing
prefactor\cite{Lambrecht}, which is also present in our calculations.
However, in exploring the continuous frequency spectrum in higher
dimensions, this single point is easily bypassed, and there is a 
divergence for all frequencies satisfying
$\omega^2/c^2 > q^2+\pi^2/H^2$, where the inequality holds in its  
strict sense. 

Resonant dissipation has profound consequences for motion of 
plates. It implies that due to quantum fluctuations of vacuum, 
{\it components of motion with frequencies in the range of 
divergences cannot be generated by any finite external force}! 
The imaginary parts of the kernels are proportional to the total 
number of excited photons \cite{Lambrecht}. Exciting these degrees
of motion must be accompanied by the generation of an infinite 
number of photons; requiring an infinite amount of  energy, and thus
impossible. However, as pointed out in Ref.\cite{Lambrecht}, the 
divergence is rounded off by assuming finite reflectivity and 
transmissivity for the mirrors. Hence, in practice, the restriction is 
softened and controlled by the degree of ideality of  the mirrors in 
the frequency region of interest. 

We shall next examine the constant term in Eq.(\ref{fi0w}).
For two plates corrugated at the same wavelength, with
deformations $h_1({\bf x})=d_1 \cos({\bf k}\cdot {\bf x})$ and
$h_2({\bf x})=d_2 \cos({\bf k}\cdot {\bf x}+\alpha)$, there is a 
(time independent) lateral force 
\begin{equation}
{\bf F}_{dc}=\frac{\hbar c A}{2} A_{-}(k,0) {\bf k} d_1 d_2 
	\sin{\alpha},
\label{JosephsonDC} 
\end{equation}
which tends to keep the plates 180 degrees out of phase, i.e.  
mirror symmetric with respect to their mid-plane.
The dependence on the sine of the phase mismatch is
reminiscent of to the DC Josephson current in superconductor  
junctions, the force playing a role analogous to the current 
in SIS junctions. There is also an analog for the AC Josephson
effect, with velocity (the variable conjugate to force) playing 
the role of voltage: Consider two corrugated plates separated 
at a distance $H$, described by
$h_1({\bf x},t)=d_1 \; \cos[{\bf k} \cdot ({\bf x}-{\bf r}(t))]$ and
$h_2({\bf x},t)=d_2 \; \cos[{\bf k} \cdot {\bf x}]$.
The resulting force at a constant velocity (${\bf r}(t)= {\bf v}\;t$),
\begin{equation}
{\bf F}_{ac}=\frac{\hbar c A}{2} A_{-}(k,0) {\bf k} d_1 d_2 
	\sin[({\bf k} \cdot {\bf v})\; t],
\label{JosephsonAC} 
\end{equation}
oscillates at a frequency $\omega={\bf k} \cdot {\bf v}$. 
Actually both effects are a consequence of the attractive
nature of the Casimir force. It would be difficult to separate them
from similar forces resulting from say, van der Waals attractions.

As a final example, we study the capillary waves on the surface
of mercury, with a conducting plate placed at a separation $H$
above the surface . The low frequency--wavevector expansion of the 
kernel due to quantum fluctuations in the intervening vacuum, starts
with quadratic forms $q^2$ and $\omega^2$. These terms result in 
corrections to the (surface) mass density by 
$\delta \rho=\hbar B/48 c H^3$, and to the surface
tension by $\delta \sigma=\hbar c B/48 H^3$. The latter correction
is larger by a factor of $(c /c_s)^2$, and changes the
velocity $c_s$, of capillary waves by 
$\delta c_s/c_s^{0}=\hbar c B/96 \sigma H^3$,
where $\sigma$ is the bare surface tension of mercury. 
Taking $H\sim 1 $mm and $\sigma \sim 500$ dynes/cm, we find 
another very samll correction of $\delta c_s/c_s^{0} \sim 10^{-19}$.

In summary, we have developed a path integral formulation for the
study of quantum fluctuations in a cavity with dynamically deforming  
boundaries. As opposed to previous emphasis on spectra of emitted
radiation, we focus on the mechanical response of the vacuum.
Most of the predicted phenomena, while quite intriguing theoretically,
appear to be beyond the reach of current experiment: the most
promising candidates are the anisotropy in mass, and resonant 
dissipation. The path integral method is quite versatile, and future
extensions could focus on non--linear response, other geometries
(e.g. wires), the gauged electromagnetic field, and calculations of
emitted spectra using correlation functions. 

RG acknowledges many helpful discussions with M.R.H. Khajehpour, 
B. Mashhoon, S. Randjbar-Daemi, and Y. Sobouti, and support from  
the Institute for Advanced Studies in Basic Sciences, Gava Zang, 
Zanjan, Iran.  MK is supported by the NSF grant DMR-93-03667.

\begin{figure}
\epsfysize=2.1truein
\epsffile{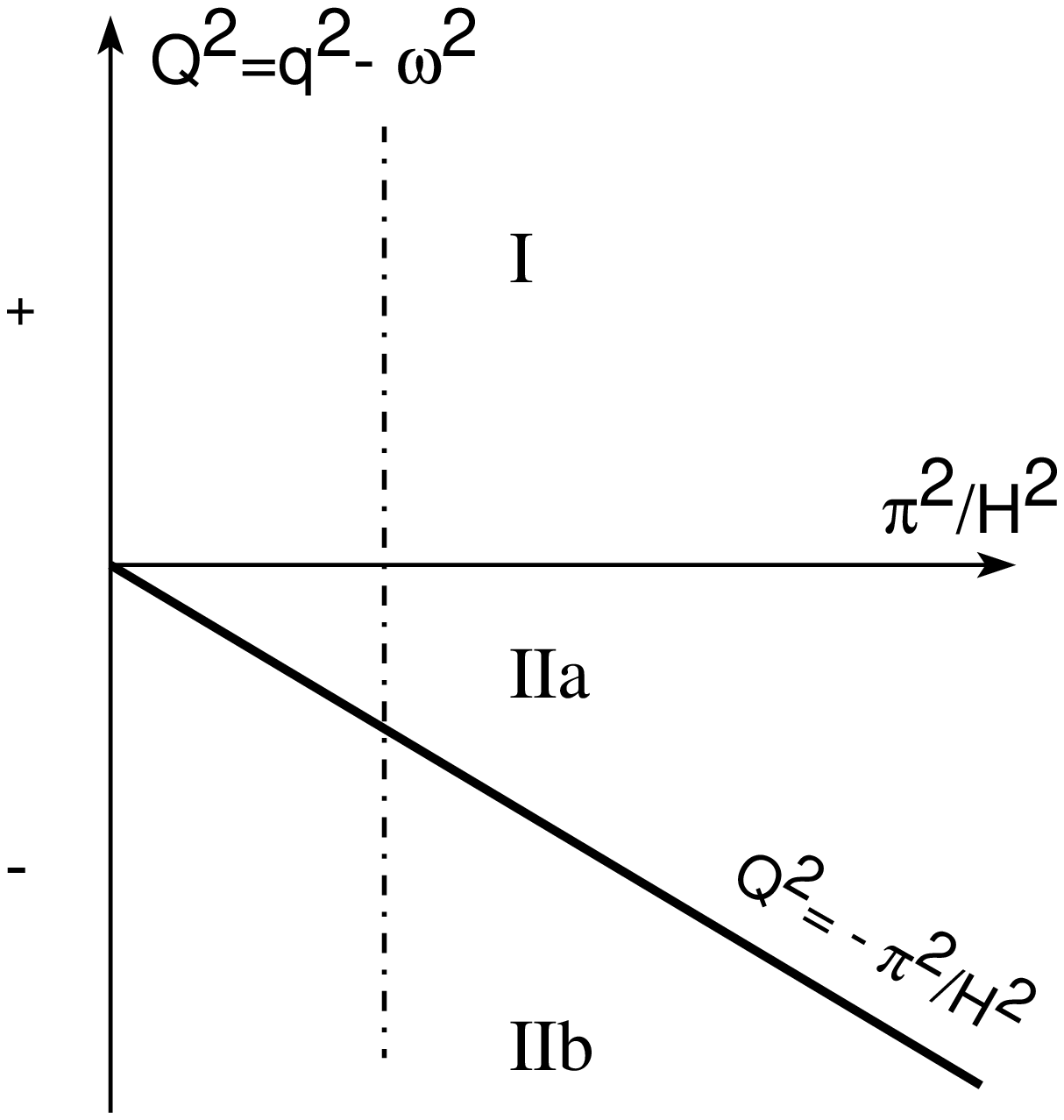}
Fig.1 Different regions of the $({\bf q},\omega)$ plane.
\label{Fig1}
\end{figure}

\end{multicols}


\begin{references}
\bibitem{Ford}
It is possible that our results can also be obtained using the 
approach of L.H. Ford and A. Vilenkin, Phys. Rev. D {\bf 25},
2569 (1982); although we find the treatment of the boundary
conditions more cumbersome.
\bibitem{EM}
Calculations with the electromagnetic field are complicated by
the requirement of gauge fixing. However,  the final results 
only change by a numerical prefactor. We have explicitly
checked that we reproduce the known answer for flat plates 
by this method.
\bibitem{cutoff}
The divergence of kernels in IIb comes from integrations over
space-time. Given a cut-off $L$ in plate size, and an associated
cutoff $L/c$ in time, the kernels diverge as 
$\exp[(K-2)L/H]/[K(L/H)^3]$, with $K=2 Q H/\pi$. Some 
care is necessary in the order of limits for $(L,H)\to\infty$.
\end{references}
\end{document}